# *The Influence of UX Design on User Retention and Conversion Rates in Mobile Apps*

Aaditya Shankar Majumder
*RV University*
*School of Design and innovation*
Bangalore, India
amaju002@gold.ac.uk

*Abstract*— This paper explores the profound impact of User Experience (UX) design on user retention and conversion rates in mobile applications. As the mobile app market becomes increasingly competitive, understanding how UX design can enhance user satisfaction, engagement, and loyalty is crucial for developers and businesses. Through a comprehensive review of existing literature and statistical insights, this study identifies key UX design principles that contribute to improved user retention and conversion rates. Intuitive navigation, appealing visuals, performance optimization, and integration of user feedback emerge as essential components of effective UX design that drive app success. Applications that prioritize these elements foster a positive user experience, leading to higher engagement and greater retention. Additionally, UX design strategies, such as personalization and customization, have been shown to significantly increase conversion rates, demonstrating the critical role that tailored experiences play in app success. By analyzing these principles and their impact, this paper provides valuable insights for developers aiming to enhance user satisfaction, optimize app performance, and ultimately improve business outcomes.

## I. INTRODUCTION

Apps for mobile devices have flourished over the past ten years. The app option pool has grown to such a point, with an excess of all apps on iOS and Android platforms (Statista, 2023). This superabundance of apps has led to fierce competition between developers, forcing them to devise something not only functional but also gives users an unforgettable user experience, and this has become a primary differentiating factor today. As people become more demanding, the spotlight has been thrust towards the quality of UX design, which influences user behaviour, ultimately, app retention, and, more importantly, conversion metrics (Hassenzahl, 2010).

User retention, in general, measures how long the user sustains engaging with the application. It shows the app's value in users' eyes, leading them time and again to return (Anderson & Rainie, 2014). High retention rates generally indicate sustainable user interest, which in turn leads to increased customer loyalty and long-term business sustenance (Google, 2018). Conversion rates, however, indicate the portion of users who bring about successful outcomes such as purchases, premiums, numerous product accounting registers, etc. (Kern, 2016). These rates become critical in revealing the revenue potential of an application in driving and achieving definite objective outcomes by using such applications.

Greater retention and conversion rates occur when an app is well and carefully designed, particularly to meet people's expectations and subtract any likely hurdles from the user journey (Shneiderman & Plaisant, 2010). To this end, the orientation of good website design-indicative, attractive display and aesthetics, performance optimization, and the seamless integration of feedback lock in the users for good.

## II. LITERATURE REVIEW

For quite some time now, this User Experience (UX) design has become one of the top priorities in the world of mobile application design. This has drawn the most attention particularly because of the way it has affected sticking around with the final users, which in the long run has been reflected in increased conversion rates. At the same time, many have found UX design to be that factor that significantly affects how users interact with mobile apps and contributes to a user's willingness to carry on using the apps. Hence this research tried to look into the established literature and their works about the correlates of transfer effects on such Key Performance Indicators in terms of user satisfaction, retention, conversion, acquisition, etc.

### 1. User Retention Rate and its Relationship with UX Design

User retention is highly acknowledged as one of the main success indicators within processes concerning an app. Retention, as defined by Anderson and Narus (2017), refers to the ability of an app to engage itself with a user over a given amount of time. Well-designed user experiences can directly work toward retaining users by ensuring a fun, smooth, and frictionless time spent with an app. A study by Forrester Research (2016) indicated that well-designed user experiences carry people through the app longer than expected, with research suggesting that a 1% decrease in customer satisfaction leads to a 10% reduction in retention rates.

There are significant insights to improve the understanding of retention by obtaining user-oriented cues that contribute to UI design, comprehend navigability, and harness personalization: According to Nielsen (2012), the moment when an interface or navigation is not observable, users usually give up on the application. In this endeavour, streamline the pathway of designing users and create opportunities for cognitive load reduction.

In defining personalization, it is the designing approach that caters more for certain needs with utmost importance placed on the user, as it was pointed out by Lemon and Verhoef (2016). Privacy, like changing the content or giving handy alerts, enhances the attractiveness and usefulness of what is to be retained.

**2. Conversion Rates and UX Design**

Conversion rates, defined as the percentage of users who perform a desired action, such as purchasing or subscribing to a service, are greatly affected by the UX designs themselves. Kim et al. emphasize that clear, simple, and intuitive navigation together with a dominant CTA button function to increase conversion rates. According to Forrester Research (2016), while UX design itself may improve website usability standards on the web, it could improve conversion rates by up to 200%, therefore noting an increase in rate with business outcomes.

Apart from the load time and what apps do when they freeze or crash, there may also potentially be low friction within the funnel, since quick-loading and easy-to-use apps are more likely to engage customers in their transactions than app crash hobbles.

According to Google (2018), almost 53% of users would vacate a site because they deemed it too slow-delayed, in fact, if it exceeded a three-second window.

Another essential element of converting dynamics says Chaffey, is how much reactive orientation is, UX, trust. By appearance, visual appeal of design, with familiar and consistent elements, the credibility and trust towards you by fostering users; in other words, esteemed as trustworthy which makes conversion possible or happening.

**3. The Role of Visual Design and Aesthetics in UX**

Visual design elements such as colour schemes, typography, and layout can influence user perceptions of the app's credibility and functionality. Tractinsky et al. (2000) found that users form impressions about the credibility of an app within just seconds of interacting with it, primarily based on aesthetic factors. This quick judgment underscores the importance of visually appealing interfaces in both attracting and retaining users.

Google's research (2016) further supports this, revealing that users tend to abandon apps with unappealing or inconsistent design elements. A clean, organized, and visually cohesive interface fosters a sense of trust and encourages users to explore the app further, which can positively impact both retention and conversion rates.

**4. Performance Optimization and Its Impact on User Experience**

App performance is an integral aspect of UX that directly affects both retention and conversion rates. Apptentive's (2020) study reveals that performance issues such as slow loading times, app crashes, and laggy interactions are among the most common reasons for user abandonment. On the other hand, performance optimization—such as faster load times and smoother transitions—can significantly enhance user satisfaction and retention. According to Apptentive (2020), apps that prioritize performance improvements see up to a 30% higher retention rate compared to those that neglect performance optimization.

In addition to performance, Kern & Wessel (2016) emphasize the importance of responsiveness in user interactions. A responsive app ensures that users receive immediate feedback when interacting with buttons, menus, or forms, making the experience feel more fluid and rewarding.

**5. The Impact of User Feedback on UX Design**

User feedback is essential for iterative improvement in UX design. VanderLinden et al. (2018) suggest that apps that incorporate feedback mechanisms—such as surveys, reviews, or in-app support—into their design process tend to improve user retention and conversion rates. By listening to user concerns and making necessary adjustments, developers can foster a sense of involvement and engagement, which in turn enhances the overall user experience.

Moreover, Gonzalez & Liu (2019) demonstrate that continuous updates based on user feedback can significantly improve app usability and customer satisfaction, leading to improved retention and higher conversion rates. When users see that their input is valued, they are more likely to stay engaged and continue using the app.

### III. IMPORTANCE OF UX DESIGN

UX design encompasses all aspects of a user's interaction with an application, including usability, accessibility, and overall satisfaction. It represents a holistic approach to designing applications that prioritize user needs and expectations, ensuring that interactions are seamless, efficient, and enjoyable (Hassenzahl, 2010). Effective UX design not only enhances the user journey but also directly contributes to key performance metrics such as retention, engagement, and conversion rates.

Research highlights several critical benefits of effective UX design:
- **Increased User Satisfaction**: A seamless and intuitive experience fosters higher customer satisfaction, which is essential for retaining users in a competitive app market. When users feel their needs are met effortlessly, they are more likely to develop loyalty toward the app (Nielsen & Norman, 2016).
- **Higher Engagement Levels**: Engaging and visually appealing user interfaces encourage longer usage sessions. Well-designed apps capture users' attention effectively, increasing the time spent within the application and facilitating deeper interactions with its features (Shneiderman & Plaisant, 2010).
- **Lower Bounce Rates**: Intuitive and user-friendly designs reduce bounce rates by ensuring users can navigate the app easily. When users encounter frustrating or complex interactions, they are more likely to abandon the app entirely. Effective UX design minimizes these frustrations, encouraging users to explore further (Google, 2018).

Statistical insights underline the critical relationship between UX design and business outcomes. Forrester Research (2016) reports that a well-designed user interface can increase conversion rates by up to 200%. Similarly, Adobe's UX Impact Study (2020) found that companies prioritizing UX design see a significant improvement in customer retention and overall revenue. These findings demonstrate the direct correlation between UX design excellence and business success, making UX design an indispensable aspect of mobile application development.

## IV. KEY ELEMENTS OF EFFECTIVE UX DESIGN

### 3.1 Intuitive Navigation

Intuitive navigation is a cornerstone of effective UX design, ensuring that users can effortlessly locate and interact with features within an application. Poor navigation is one of the primary reasons users abandon apps; in fact, research indicates that 79% of users stop using an app due to confusing or cumbersome navigation (Google, 2018). Clear menus, logical flow, and easily accessible features are critical for creating a positive user experience and retaining users.

Case Study: E-Commerce App

In the context of e-commerce applications, intuitive navigation plays a pivotal role in the user journey. For instance, a strategically placed and visible "checkout" button can significantly simplify the purchasing process and reduce cart abandonment rates (Baymard Institute, 2021). By optimizing navigation pathways—such as integrating breadcrumb trails or categorizing products logically—developers can enhance user satisfaction, making it easier for customers to complete transactions. This not only improves retention but also drives conversions, directly impacting revenue generation.

Furthermore, apps with well-designed navigation often utilize features like search bars, filters, and predictive text to improve usability. These elements help reduce user frustration by minimizing the time and effort required to perform specific tasks, thus increasing engagement and loyalty (Nielsen & Norman, 2016).

### 3.2 Visual Appeal

Visual design is a critical factor in attracting and retaining users, as aesthetic elements such as color schemes, typography, and layout strongly influence user perceptions and trust in an app. Research has shown that users form judgments about an app's credibility based on its visual design within seconds (Fogg et al., 2003). A clean and cohesive design can enhance usability and establish a positive emotional connection with users, fostering loyalty and engagement.

**Statistical Insight**

A study conducted by Google revealed that users form an impression of an app within just 50 milliseconds (Google, 2012). This highlights the importance of visually appealing designs in creating positive first impressions. Developers who prioritize aesthetics, while maintaining usability, are more likely to capture and sustain user interest, ultimately driving higher retention and conversion rates.

### 3.3 Performance Optimization

App performance is a cornerstone of user experience, as slow loading times, glitches, or crashes can lead to frustration and abandonment. Ensuring smooth and reliable app functionality is essential for maintaining user engagement and satisfaction. Performance optimization includes reducing load times, improving responsiveness, and ensuring compatibility across devices.

**Statistical Insight**

Research from Apptentive indicates that apps with performance optimizations see a 30% higher retention rate, underscoring the significant impact of technical reliability on user loyalty (Apptentive, 2021). Users expect seamless interactions, and even minor delays can negatively affect their perception of the app. For example, Amazon found that a 100-millisecond delay in load time led to a 1% drop in sales, demonstrating the broader implications of app performance on business outcomes (Kohavi et al., 2007).

By prioritizing performance optimizations alongside design elements, developers can create an application that not only attracts users but also ensures sustained engagement and satisfaction over time.

## V. IMPACT ON USER RETENTION

User retention is one of the most critical metrics for evaluating the success of a mobile application. It reflects the app's ability to maintain a loyal user base and sustain engagement over time. Retaining users is often more cost-effective than acquiring new ones, making it an essential focus for developers and businesses. Several factors significantly influence retention:

- **User Satisfaction**: A positive user experience fosters loyalty, encouraging users to return consistently. Apps that meet or exceed user expectations through seamless navigation, attractive design, and reliable performance are more likely to retain users (Nielsen & Norman, 2016).
- **Engagement Metrics**: Apps that effectively engage users—through gamification, personalized content, or push notifications—experience higher retention rates. Engaged users are more likely to integrate the app into their daily lives, ensuring long-term usage (Anderson & Rainie, 2014).
- **Feedback Mechanisms**: Incorporating user feedback into app updates demonstrates responsiveness and a commitment to improvement. This practice not only enhances user satisfaction but also increases retention by addressing pain points and evolving with user needs (Apptentive, 2021).

**Statistical Insight**

Research underscores the critical importance of making a strong first impression. For instance, a well-designed onboarding process can increase user retention by up to 50% (Google, 2018). Onboarding processes that demonstrate an app's value, guide users through key features and provide an intuitive introduction to functionality can create a positive initial experience, motivating users to stay engaged with the app.

By focusing on these factors, developers can foster greater user loyalty, minimize churn, and establish a sustainable user base that contributes to the app's long-term success.

## VI. INFLUENCE ON CONVERSION RATES

Conversion rates are vital for monetizing mobile applications, as they measure the percentage of users who perform desired actions, such as making purchases, subscribing to services, or completing registrations. Effective UX design plays a pivotal role in enhancing conversion rates by streamlining user interactions and building trust. Key factors that influence conversion through UX design include:

- **Guided User Journeys**: Creating clear and intuitive pathways for completing tasks ensures that users can easily navigate the app to achieve their goals. Features like well-placed call-to-action buttons, logical page flow, and progress indicators reduce friction in the user journey, increasing the likelihood of conversions (Nielsen & Norman, 2016).
- **Trust Building**: Professional and visually appealing design fosters trust and credibility, which are critical for users to feel comfortable engaging in transactions. Elements like consistent branding, secure payment gateways, and clear communication of app policies instil confidence in users (Fogg et al., 2003).
- **Error Reduction**: Minimizing user errors during interactions, such as form submissions or checkout processes, leads to smoother experiences and higher conversion success rates. Techniques like real-time form validation and auto-filling fields can reduce frustration and ensure a seamless process (Shneiderman & Plaisant, 2010).

**Statistical Insight**

According to research from AI Marketing Engineers, effective UX design can potentially raise customer conversion rates by up to 400% (AI Marketing Engineers, 2021). This statistic underscores the transformative impact of prioritizing UX in app development strategies. By focusing on user-centric design principles, developers can create apps that not only attract users but also guide them toward meaningful actions, driving business success.

Investing in UX design is, therefore, not merely an aesthetic choice but a strategic approach to achieving higher monetization and user satisfaction.

## VII. REAL-WORLD EXAMPLES

### 6.1 Case Study: Mobile Shopping App

A mobile shopping application struggled with high churn rates due to a complicated onboarding process and inconsistent design elements, such as misaligned typography, unclear navigation paths, and confusing colour schemes. Users found it challenging to navigate the app or complete tasks like searching for products or checking out.

To address these issues, the app underwent a UX redesign, focusing on intuitive navigation, streamlined onboarding, and clear calls-to-action. Enhancements included simplifying the registration process, creating a more organized product catalogue, and integrating a prominent and easy-to-find "checkout" button. As a result:

- **Retention rates** increased by 30%, as users were more likely to return and explore the app.
- **Conversion rates** improved by 200%, demonstrating the power of UX in influencing purchasing behaviour (Forrester Research, 2016).

This case study highlights the importance of aligning app design with user expectations and providing frictionless user journeys.

**6.2 Case Study: Fitness Tracking App**

A fitness tracking app sought to enhance user engagement and retention by leveraging data-driven personalization. The app integrated features such as personalized workout recommendations, tailored fitness goals, and adaptive notifications based on user activity.

These changes created a more engaging and relevant user experience, making users feel that their unique needs and preferences were being addressed. Over six months, the app observed:

- A 25% increase in retention rates, as users found value in the customized recommendations.
- Enhanced engagement metrics, with users spending more time tracking their progress and achieving fitness milestones.

This case demonstrates how personalization, a core element of UX design, can significantly impact user satisfaction and long-term loyalty (Adobe UX Impact Study, 2020).

## VIII. CONCLUSION

In conclusion, prioritizing high-quality UX design is no longer an option but a necessity for mobile applications aiming to enhance user retention and conversion rates. By incorporating key elements such as intuitive navigation, visually appealing design, performance optimization, and a strong feedback integration process, developers can craft experiences that not only attract users but also retain them over time.

Effective UX design creates a seamless journey for users, reducing friction and fostering trust, which directly contributes to increased engagement, loyalty, and monetization opportunities. The real-world examples discussed in this paper further highlight the tangible benefits of implementing user-centric design principles in diverse app categories.

As the mobile app market becomes increasingly competitive, UX design will serve as a critical differentiator for applications striving for long-term success. Apps that prioritize user needs, adapt to changing expectations, and employ innovative design strategies will be better positioned to thrive in this dynamic environment.

Future research should delve deeper into emerging UX trends, such as artificial intelligence-driven personalization, augmented reality interfaces, and adaptive design systems, to further understand their impact on user engagement and satisfaction. By continuing to innovate in UX design, developers and businesses can unlock new opportunities for creating meaningful and impactful user experiences.

## X. ACKNOWLEDGMENT

I would like to express my deepest gratitude to all those who supported and guided me throughout the process of completing this research.

First and foremost, I extend my heartfelt thanks to my advisor, Nick Hine, whose expertise, encouragement, and constructive feedback have been invaluable in shaping this study. Your insights and unwavering support have been a constant source of inspiration.

I am profoundly grateful to the participants who generously contributed their time and experiences to this research. Your input provided the foundation for this study and enriched its outcomes.

Special thanks go to my colleagues and peers who provided thoughtful discussions, suggestions, and camaraderie during this journey. Your collaboration and shared passion for this field have been instrumental in refining my ideas.

Finally, I am deeply indebted to my family and friends for their unconditional support and encouragement. Your faith in me has been a source of strength and motivation, allowing me to pursue these endeavours with confidence.

This work would not have been possible without the collective efforts of all those mentioned above. Thank you for helping bring this vision to life.